
\input phyzzx

\batchmode\font\eightcp=cmcsc8\errorstopmode
\ifx\eightcp\nullfont\font\eightcp=cmr8\fi
\def\runningheads#1{\paperheadline={\iffrontpage\else
	{\hfil{\eightcp #1}\hfil}\fi}}

\def\a{\alpha}
\def\b{\beta}
\def\d{\delta}
\def\e{\eta}
\def\m{\mu}
\def\n{\nu}
\def\l{\lambda}
\def\th{\vartheta}
\def\o{\omega}
\def\p{\pi}
\def\x{\xi}
\def\g{\gamma}

\def\P{\Pi}

\def\punkt{\hskip2mm .}
\def\komma{\hskip2mm ,}
\def\is{\!=\!}

\def\frac#1/#2{\leavevmode\kern.1em\raise.5ex
		\hbox{\the\scriptfont0\def\nicefrac#1#2{\hbox{${#1\over #2}$}}
         	#1}\kern-.1em/\kern-.15em
		\lower.25ex\hbox{\the\scriptfont0 #2}}

\def\nicefrac#1#2{\hbox{${#1\over #2}$}}
\def\half{\nicefrac 12 }

\runningheads{M.~Cederwall, ``A Note on the Relation between Different
	Forms of Superparticle Dynamics''}
\date{October, 1993}
\pubnum={\vbox{	\hbox{G\"oteborg-ITP-93-33}
		\hbox{hep-th/9310177}}}
\titlepage
\title{\fourteenpoint A Note on the Relation between Different\break\break
	Forms of Superparticle Dynamics}
\author{Martin Cederwall}
\address{Institute for Theoretical Physics\break
	Chalmers University of Technology and University of G\"oteborg\break
	S-412 96 G\"oteborg, Sweden}
\vskip1cm
\abstract
A formulation of $D\is 10$ superparticle dynamics is given that
contain space-time and twistor variables. The set of constraints is
entirely first class, and gauge conditions may be imposed that reduces
the system to a Casalbuoni-Brink-Schwarz superparticle, a spinning particle or
a twistor particle.

\endpage

\REF\STVZ{D.P.~Sorokin, V.I.~Tkach, D.V.~Volkov and A.A.~Zheltukhin,
	\sl Phys.Lett. \bf 216B \rm (1989) 302.}
\REF\CBS{R. Casalbuoni, \sl Nuovo Cim. \bf 33A \rm (1976) 389;\nextline
	L.~Brink and J.H.~Schwarz, \sl Phys.Lett. \bf 100B \rm (1981) 310.}
\REF\Spinn{L.~Brink, S.~Deser, B.~Zumino, P.~Di Vecchia and P.~Howe,
	\sl Phys.Lett. \bf 64B \rm (1976) 435.}
\REF\kappasiegel{W. Siegel, \sl Phys.Lett. \bf 128B \rm (1983) 397,
	\sl Class.Quant.Grav. \bf 2 \rm (1985) 195.}
\REF\IBMC{I. Bengtsson and M. Cederwall, G\"oteborg-ITP-84-21 (1984).}
\REF\SSpart{D.P.~Sorokin, V.I.~Tkach and D.V.~Volkov,
	\sl Mod.Phys.Lett. \bf A4 \rm (1989) 901;\nextline
	F.~Delduc and E.~Sokatchev,
	\sl Class.Quant.Grav. \bf 9 \rm (1992) 361.}
\REF\multitwist{A.~Galperin, P.S.~Howe and K.S.~Stelle,
	\sl Nucl.Phys. \bf B368 \rm (1992) 248;
	\nextline A.~Galperin and E.~Sokatchev,
	\sl Phys.Rev. \bf D46 \rm (1992) 714.}
\REF\DivTwist{I. Bengtsson \sl Class.Quantum Grav. \bf 4 \rm (1987) 1143;
	\nextline A.K.H. Bengtsson, I. Bengtsson, M. Cederwall and N. Linden,
	\sl Phys.Rev. \bf D36 \rm (1987) 1766;\nextline
	I.~Bengtsson and M.~Cederwall, \sl Nucl.Phys. \bf B302 \rm (1988) 81;
	\nextline M.~Cederwall, \sl Phys.Lett. \bf 210B \rm (1988) 169.}
\REF\tentwistor{N.~Berkovits, \sl Phys.Lett. \bf 247B \rm (1990) 45;\nextline
	M.~Cederwall \sl J.Math.Phys. \bf 33 \rm (1992) 388.}
\REF\ESTPS{F. Englert, A. Sevrin, W. Troost, A. Van Proyen and Ph. Spindel,
	\nextline\indent\sl J.Math.Phys \bf 29 \rm (1988) 281.}
\REF\sevensphere{M.~Cederwall and C.R.~Preitschopf,
	G\"oteborg-ITP-93-34, hep-th/9309030}
\REF\Eightconf{L.~Brink, M.~Cederwall and C.R.~Preitschopf,
	\sl Phys.Lett. \bf B311 \rm (1993) 76.}
\REF\EisenbergParticle{Y.~Eisenberg and S.~Solomon,
	\sl Nucl.Phys. \bf B309 \rm (1988) 709.}
\REF\multi{M.~Cederwall, G\"oteborg-ITP-91-26.}
\REF\harmonic{A.~Galperin, E.~Ivanov, S.~Kalizin, V.~Ogievetsky
	and E.~Sokatchev,\nextline\indent
	\sl Class.Quant.Grav. \bf 1 \rm (1984) 469,
	\sl Class.Quant.Grav. \bf 2 \rm (1985) 155;
	\nextline E.~Sokatchev, \sl Phys.Lett. \bf 169B \rm (1986) 209;
	\nextline E.~Nissimov, S.~Pacheva and S.~Solomon,\nextline\indent
	\sl Nucl.Phys. \bf B296 \rm (1988) 462,
	\sl Nucl.Phys. \bf B297 \rm (1988) 349.}

In reference [\STVZ], Sorokin, Tkach, Volkov and Zheltukhin established
the equivalence between the Casalbuoni-Brink-Schwarz (CBS) superparticle
[\CBS] and the spinning particle in $D\is 10$ [\Spinn].
Their formulation, that has
led to progress in the understanding of the fermionic symmetries
[\kappasiegel,\IBMC ]
of supersymmetric particles [\SSpart,\multitwist],
introduced a ``twistor-like''
variable $\l^\a$, a bosonic spinor. In this note, I will show that
by making this spinor dynamical one can formulate a dynamical system
that contains the CBS superparticle and the spinning particle
as well as the division algebra twistor formulation of the
superparticle [\DivTwist,\tentwistor] as special
gauge choices. Unlike in reference [\STVZ], there is no manifest
world-line supersymmetry. It is not impossible, though, that this
constraint structure will arise from a manifest $N\is 8$ world-line
superconformally symmetric treatment based on $S^7$
[\ESTPS,\sevensphere,\Eightconf].
The present framework is rather reminiscent of
that in reference [\EisenbergParticle].

The literature on the supersymmetric point particle and its
(covariant) first quantization is vast. The list of references
only contains items of immediate interest for the issues addresses
here, and I apologize for leaving out many important contributions.
 The content of this
note is by no means revolutionary, but it is probably the
simplest way of establishing the equivalence of the different
forms of superparticle dynamics.

The phase space variables are ($X^\m,P^\m$), parametrizing Minkowski
phase space, ($\th^\a,p_\a$), the fermionic variables (spinors) of
the CBS superparticle, ($\l^\a,\o_\a$), the twistor
variables, and $\x^\m$, the fermionic vector of the twistor particle
or spinning particle. Their Poisson brackets are
$$\eqalign{&\hbox to 4cm{$\{X^\m,P^\n\}=\e^{\m\n}\komma$\hfill}
	\hbox to 4cm{$\{\th^\a,p_\b\}=\d^\a_\b\komma$\hfill}\cr
	&\hbox to 4cm{$\{\l^\a,\o_\b\}=\d^\a_\b\komma$\hfill}
	\hbox to 4cm{$\{\x^\m,\x^\n\}=-\e^{\m\n}\punkt$\hfill}\cr}
						\eqn\phasePBs$$
For this set of variables, I postulate the constraints
$$\eqalign{&\P^\m\equiv P^\m-\half(\l\g^\m\l)\approx 0\komma\cr
	&\p_\a\equiv p_{\a}+P_\m(\g^\m\th)_\a+\x_\m(\g^\m\l)_\a\approx 0
								\komma\cr
	&T^\a\equiv\half(\l\g_\m\l)(\g^\m\o)^\a-\l^\a(\l\o)
		-\half \x_\m\x_\n(\g^{\m\n}\l)^\a\approx 0\komma\cr
	&t\equiv\half(\l\g_\m\l)\x^\m\approx 0\punkt\cr}\eqn\constraints$$
The first constraint states the twistor transformation of the
particle momentum, and implies its lightlikeness.
The second one is the usual set of fermionic constraints for the
superparticle, modified with the last term which makes it first class.
The last two constraints are the $S^7$ (Hopf map) generators
[\tentwistor,\sevensphere] of the twistor
string and its fermionic companion. They have a second order (covariant)
bosonic reducibility, that reduces the number of independent $T$'s to $7$.
The whole set of constraints can be seen as defining the ``twistor transform''
between the different forms of superparticle dynamics.

The non-vanishing Poisson brackets between constraints are
$$\eqalign{&\{\p_\a,\p_\b\}=2\g_{\m\a\b}\P^\m\komma\cr
	&\{T^\a,\p_\b\}=-2\d^\a_\b t\komma\cr
	&\{T^\a,T^\b\}=-\l^\a T^\b+\l^\b T^\a
	    =\nicefrac{1}{48}\g_{\m\n\rho}^{\a\b}(\l\g^{\m\n\rho}T)\komma\cr
	&\{T^\a,t\}=-\l^\a t\punkt\cr}\eqn\constraintPBs$$
and one notices that $\P$ and $\p$ generate an ordinary $D\is 10$ supersymmetry
algebra, that has now become a gauge symmetry.

It is now straightforward to partially fix the gauge in different ways
to arrive at the various advertised formulations:\endpage

\item{i.}Use $\P$ to eliminate ($X,P$) and $\p$ to eliminate ($\th,p$).
	This gives the twistor formulation.

\item{ii.}Use $T$ and $\P$ (except $P^2\is 0$) to eliminate ($\l,\o$)
	and $t$ and part of $\p$ to eliminate $\x$. This gives
	the CBS superparticle (the fermionic second class
	constraints arise via Dirac brackets when the whole of $\p$
	is retained).

\item{iii.}Use $T$ and $\P$ (except $P^2\is 0$) to eliminate ($\l,\o$)
	and $\p$ to eliminate ($\th,p$). This gives the
	spinning particle.

There is of course classes of formulations not reached through this
procedure. These are the models of reference [\multitwist], where
the twistor variables are multicomponent [\multi], and not of
the simple division algebra type, and the ones using harmonic
variables [\harmonic].

It is likely that this work has a connection to a not yet formulated
$N\is 8$ supersymmetric formalism based on $S^7$. It is also
plausible that extended supersymmetric field theories in $D\is 10$
may be given an off-shell symmetric formulation using the set of
variables of this note.

\refout

\end